\documentclass[twocolumn,showpacs,prb,amsmath,amssymb,superscriptaddress]{revtex4}
\usepackage{graphicx}% Include figure files
\usepackage{array}
\begin{document}

\title{Excitons in layered metal halide perovskites: an effective mass description of polaronic, dielectric and quantum confinement effects}

\author{Jose L. Movilla}
\affiliation{Dept. d'Educaci\'o i Did\`actiques Espec\'ifiques, Universitat Jaume I, 12080, Castell\'o, Spain}

\author{Josep Planelles}
\affiliation{Dept. de Qu\'imica F\'isica i Anal\'itica, Universitat Jaume I, 12080, Castell\'o, Spain}

\author{Juan I. Climente}
\email{climente@uji.es}
\affiliation{Dept. de Qu\'imica F\'isica i Anal\'itica, Universitat Jaume I, 12080, Castell\'o, Spain}
\homepage{http://quimicaquantica.uji.es/}

\date{\today}

\begin{abstract}

A theoretical model for excitons confined in layered metal halide perovskites is presented.
The model accounts for polaronic effects, dielectric and quantum confinement
by means of effective mass theory, image charges and Haken potentials.
We use it to describe the band edge exciton of MAPbI$_3$ structures surrounded by organic ligands.
 It is shown that the quasi-2D quantum and dielectric confinement of layered perovskites
squeezes the exciton radius, and this in turn enhances polaronic effects as compared to 3D structures.
 The strong polaronic effects boost the binding energies and radiative recombination probabilities,
which allows one to match experimental data in related systems. 
 The thickness dependence of Coulomb polarization and self-energy potentials 
 is in fair agreement with sophisticated atomistic models.
\end{abstract}

\maketitle

%%%MAIN TEXT%%%%

\section{\label{secintro} Introduction}

In the last years, 2D layered hybrid (organic-inorganic) metal halide perovskites have emerged as a new class of quasi-two-dimensional materials with outstanding optical (photovoltaic, light-emitting) properties and improved moisture stability as compared to their 3D counterparts.\cite{SmithANG, LantyJPCL, TsaiNAT, HuJPCL, MaNs, PedesseauACS, LiuACS, TsaiAM, SoeAEM, WangAOM, ZhouAMI, LiAEM}
The optical properties of these objects (also referred to as nanoplatelets, NPLs) are intimately connected with the electronic structure of the band edge exciton.
A large number of experimental and theoretical studies have investigated how the exciton electronic structure changes with the thickness and composition of either the inorganic layers, and those of the organic ligands in between consecutive layers.\cite{LiuACS, ChakraJPCc, TraoreACS, ZhangJPCL, YinACS, PassarelliNC, TanAEM, EvenCPC, SmithCS, BlanconNC, TanakaPRB, SaporiNS, GhribiNm, BohnNL}
It is now clear that the electronic structure results from an interplay between the strong quantum confinement of the inorganic layers (the ligands have large gaps and hence act as potential barriers) and the strong dielectric confinement (the ligands have typically small dielectric constant).

While the separation between quantum and dielectric confinement effects is difficult by experimental means, computational simulations allow one to visualize the impact of each factor,\cite{ChakraPCCP} thus paving the way towards optimal engineering of optoelectronic devices based on these materials.\cite{GengATS}
A similar scenario has been found in II-VI NPLs. Here, effective mass models have proved very useful in providing intuitive interpretations of a wide variety of experimental observations. These include exciton energy and oscillator strength dependence on the composition and material thickness,\cite{IthurriaNM, AchtsteinNL, ChristodoulouNL, BenchamekhPRB, DabardCM} exciton binding energies,\cite{RajadellPRB, ShornikovaNL, ZelewskiJPCL, YangJPCc, AyariNS, MaciasJPCc} heterostructure details,\cite{RajadellPRB, JordiJPCc, KhanACS, DabardNC, PolovitsynCM} absorption and emission directionality\cite{ScottNN} and non-linear optical properties.\cite{HeckmannNL, PlanellesAP}
It is naturally desirable to extend effective mass models to halide perovskite quasi-2D materials. Input parameters such as effective masses and dielectric constants can be inferred from atomistic simulations already present in the literature.\cite{KatanCR}

Initial studies have been already conducted in this direction.\cite{GhribiNm} However, there is a physical factor which has not been considered to date, and yet it constitutes a differential trait with respect to II-VI NPLs. Namely, the presence of sizable polaronic effects.\cite{BaranowskiAEM}
Polaron effects entail non-negligible carrier-phonon coupling. These are significant in hybrid metal halide perovskites because of the softness of the lattice,\cite{FerreiraPRL} the low phonon energies\cite{LetoublonJPCL, SendnerMH, LeguyPCCP} and the high difference between static and high-frequency dielectric constants in these materials.\cite{SendnerMH, AnuscaAEM, MelissenPCCP, WilsonAPLM} All these factors lead to Fr\"{o}lich constants (exciton-LO phonon coupling) up to one order of magnitude greater than in II-VI materials.\cite{BaranowskiAEM}

There has been increasing awareness of the importance of polaron effects not only in 3D bulk hybrid perovskites,\cite{ThouinNM, WolfCSC, BaoPRXe, MenendezPSS} but also in 2D layered ones. They have been shown to increase exciton effective masses,\cite{BaranowskiAEM} affect their spin dynamics,\cite{BourelleNC, TaoSA} radiative recombination,\cite{NeutznerPRM} binding energy\cite{SichertNL} and even lead to self-trapping.\cite{Zhang2D}

Because direct inclusion of polaron effects in exciton calculations is a major challenge, semi-empirical approximations have been proposed in the literature. One such case is the Haken potential.\cite{Haken} By including the electron and hole polaron radii -which can be inferred from experiments and atomistic calculations\cite{SendnerMH, Zhang2D}- as a parameter, the electron-hole Coulomb potential is rewritten as:
\begin{equation}
\label{eq1}
	V(r)=-\frac{1}{\epsilon_{s}}\frac{q^2}{r} - \frac{1}{\epsilon^{*}}\frac{q^2}{r} (e^{-r/l_e}+e^{-r/l_h}) = V_C + V_Y.
\end{equation}
The above expression, which is explained in more detail in the next section, contains a first term ($V_C$) describing the standard Coulomb interaction (with full, static screening) plus a second term ($V_Y$) which introduces the ionic screening correction, due to carrier-lattice coupling at short distances. The latter mimics the inability of the lattice to screen the interaction when the electron-hole pair is very close (below the polaron radius), thus enhancing excitonic interactions.
The Haken potential (and similar potentials, such as the Bajaj\cite{BajajSSC} and Pullman-B\"{u}ttner\cite{PB} ones) have been shown to improve the estimates of the exciton binding energy in 3D metal halide perovskites as compared to estimates neglecting polaron effects, reconciling some controversial observations in experiments.\cite{BaranowskiAEM, MenendezPSS}

The goal of this article is to extend the effective mass theory description of excitons in quasi-2D systems, previously developed for II-VI NPLs,\cite{RajadellPRB} for the case of hybrid metal halide perovskites. This is done by considering not only quantum and dielectric confinement, but also polaron effects -through the inclusion of Haken-like potentials-.
The technical novelty lies in the conjugation of dielectric confinement and polaronic effects. The former is conveniently expressed using the image charge method for quantum well-like structures,\cite{KumagaiPRB} which has proved valuable in the description of II-VI NPLs with varying thickness.\cite{BenchamekhPRB, RajadellPRB, ShornikovaNL, PolovitsynCM, YangJPCc} Such a method was however posed for Coulomb-law potentials, $V_c$. Here we adapt it to account for the additional Yukawa-like potential present in Eq.~(\ref{eq1}), $V_Y$. 

 We derive image charge expressions for both electron-hole interaction and for the self-energy corrections. The latter result from the interactions of carriers with their own image charges. Self-energy terms, which escape from simple Keldysh potentials often used to model dielectric confinement,\cite{BlanconNC,AyariNS} are particularly important for accurate estimates of the excitonic band gap\cite{KatanCR,PolovitsynCM}
 or exciton interactions across the organic barrier separating inorganic layers.\cite{MovillaJPCL}

Our model is tested for the prototypical case of MAPbI$_3$. We find remarkable agreement with the self-energy obtained from ab-initio calculations\cite{SaporiNS} for NPLs down to a single layer, by including a single phenomenological parameter related to the non-abrupt profile of the dielectric constant in the vicinity of the organic-inorganic interfaces. We also show that polaronic interactions have a prominent role in excitons confined in few-layers metal halide perovskites. This is because the strong quantum and dielectric confinement  squeeze the exciton under the polaron radius. It is then possible to explain the large binding energies (200-500 meV) reported in experiments.\cite{BlanconNC}

\section{Theoretical Framework}
\label{sectheo}

Within the effective mass formalism, the exciton ground state of a quasi-2D structure subject to quantum and dielectric confinement 
can be obtained from:\cite{RajadellPRB}
\begin{equation}
\label{eqH}
H=\sum_{i=e,h} \left( \frac{{\mathbf p}_i^2}{2 m_i} + V_i^{self} + V_i^{pot} \right) + V_{eh} + E_{gap}^{\Gamma},
\end{equation}
\noindent where $e$ and $h$ indexes stand for electron and hole, $\mathbf{p}_i$ for the momentum operator, $m_i$ for the effective mass of 
carrier $i$ and $E_{gap}^{\Gamma}$ for the energy gap between the conduction and the valence band at the $\Gamma$ point. 
$V_i^{pot}$ is the confining potential set by the band offset between the perovskite layers and the surrounding ligands. 
Because the organic character of such ligands prevents carriers from tunneling,\cite{BlanconNC}
it is convenient to set $V_i^{pot}=0$ inside the perovskite domain and $V_i^{pot}=\infty$ outside.
$V_i^{self}$ and $V_{eh}$ are the self-energy and electron-hole Coulomb potentials, 
which can be calculated using the image-charge method for quantum wells\cite{KumagaiPRB}
to account for dielectric mismatch effects.
 
Eq.~(\ref{eqH}) has been employed on different occasions to study excitons in II-VI NPLs.\cite{RajadellPRB,ShornikovaNL,PolovitsynCM,MaciasJPCc}
Its extension to hybrid halide perovskite materials, requires including polaronic effects, 
i.e. the dressing of charge carriers by the local lattice polarization resulting from the Coulomb interaction 
between the carrier and the ionic lattice. 
This can be done by modifying $V_i^{self}$ and $V_{eh}$ with respect to their usual expressions. Below we describe the procedure to do so.

% and $\omega_{LO}$ is the optical phonon frequency. The term $(1/\epsilon_{\infty} - 1/\epsilon_s) = 1/\epsilon^*$ 
%$\left( \frac{1}{\epsilon_{\infty}}- \frac{1}{\epsilon_s}\right) = \frac{1}{\epsilon^*}$ 
%represents ionic screening of carriers, where $\epsilon_{\infty}$ and $\epsilon_s$ are optical frequency and static dielectric constants, respectively.\\

Neglecting for the time being dielectric mismatch, polaronic effects are efficiently described by the Haken potential,
which in a.u. reads:\cite{Haken}
\begin{eqnarray}
\label{eq02}
V_{eh}^{bulk}(r)&=&-\frac{q^2}{\epsilon_{\infty}\, r}+\frac{q^2}{r} \left( \frac{1}{\epsilon_{\infty}}-\frac{1}{\epsilon_{s}} \right)\, \left[1- \frac{e^{-\beta_e\, r}+e^{-\beta_h\, r}}{2}\right]\nonumber \\
\, &=& -\frac{q^2}{\epsilon_s \, r}-\frac{q^2}{\epsilon^* \,r}   \, \frac{e^{-\beta_e\, r}+e^{-\beta_h\, r}}{2}.
\end{eqnarray}
\noindent  Here, $r$ is the electron-hole distance, $q$ the fundamental charge, $\epsilon_s$ the static dielectric constant and 
$\epsilon_\infty$ the high frequency (optical) one. The term 
$\left( \frac{1}{\epsilon_{\infty}}- \frac{1}{\epsilon_s}\right) = \frac{1}{\epsilon^*}$ 
represents ionic screening of carriers.
 $\beta_i$ are the electron and hole polaron radii inverse: $\beta_i=l_{i}^{-1} = (2\, m_i\, \omega_{LO} / \hbar)^{1/2}$,
with $\hbar \omega_{LO}$ the longitudinal optical phonon frequency.

Eq.~(\ref{eq02}) can be viewed either as an increase of the Coulomb interaction by the reduction of the ionic screening of carriers at short distances (first line) or as if charges embedded in polar semiconductors had a two-fold interaction: 
a standard screened Coulomb interaction supplemented by 
a Yukawa-like interaction in a medium of effective dielectric constant $\epsilon^*$  
($V_{eh}=V_C+V_Y$, second line in the equation). 
We adopt the second point of view in this work.\\

To include dielectric mismatch in $V_{eh}$, we rely on the image charge method.
Usual expressions found in the literature are obtained from Coulomb interaction terms ($V_C$).\cite{Jackson_book,KumagaiPRB}
Here we need to revisit their derivation for the Yukawa-like terms ($V_Y$).
The Yukawa potential is not solution of the Poisson equation, 
that accounts for the electromagnetic interactions carried by massless photons, and that for a 
point source charge in vacuum reads (MKS units):
\begin{equation}
	\nabla^2 \Psi_C =-\frac{q}{\epsilon_0} \delta({\mathbf r}),
\end{equation}
\noindent with $\epsilon_0$ the vacuum permittivity and $\Psi_C$ the electrostatic potential.
Rather, the carriers of the Yukawa interactions have finite mass. 
Then, the Poisson equation should be modified by including a mass ($\mu$) term as follows:\cite{TuRPP}
\begin{equation}
\label{eq04}
\nabla^2 \Psi_Y-\mu^2 \Psi_Y=-\frac{q}{\epsilon_0} \delta({\mathbf r}),
\end{equation}

\noindent The integration of Eq.~(\ref{eq04}) yields back the Yukawa potential:
\begin{equation}
\Psi_Y(r)=\frac{q}{4 \pi \epsilon_0} \frac{e^{-\mu\, r}}{r}.
\end{equation}

In a dielectric medium other than the vacuum we should replace $\epsilon_0$ by a dielectric constant $\epsilon_1$, 
and for a given charge distribution $\rho$ we should replace $q \delta({\mathbf r})$ by $\rho$. 
All and all, Eq.~(\ref{eq04}) reads: 
\begin{equation}
\label{eq04b}
\epsilon_1 \nabla^2 \Psi_Y-\epsilon_1 \mu^2 \Psi_Y=-\rho
\end{equation}

In order to determine the interface boundary conditions between two homogeneous materials with different dielectric constant, 
and taking into account the definition of the electric displacement vector ${\mathbf D}=-\epsilon \nabla \Psi_Y$, 
we can rewrite Eq.~(\ref{eq04b}) as $\nabla \cdot {\mathbf D}=\rho- \mu'^2 \Psi_Y$. 
Then, we consider a thin, $dz$ height slab around the interface and integrate this equation,
\begin{equation}
\label{eq05}
 \int \nabla \cdot {\mathbf D} \; dV = \int \rho \,dV- \mu'^2 \int \Psi_Y dV.
\end{equation}

By taking into account the continuity of $\Psi_Y$ across the boundary, we see that as  $dz \to 0$ the second integral approaches zero. If we additionally use the divergence theorem to transform volumetric integrals into surface ones, we find:
\begin{equation}
\label{eq06}
 \int ({\mathbf D_1}-{\mathbf D_2})\cdot {\mathbf u}_z \; dA = \int \sigma \,dA  \to {\mathbf D_1}_{\perp}-{\mathbf D_2}_{\perp} = \sigma,
\end{equation}
\noindent with $\mathbf{u}_z$ a unit vector crossing the interace, $A$ the area of integration, and $\sigma$ the electric
charge density on the interface. The relevant result here is that the obtained boundary condition of $\Psi_Y$ 
coincides with the usual one for Coulomb potentials, $\Psi_C$.\\

We can also show that the charge density generated by the Yukawa field on a dielectrically mismatched interface is similar to that of the Coulomb one. The potential 
\begin{equation}
\Psi_Y(r)= \frac{1}{4 \pi \epsilon_0}\frac{1}{\epsilon_1^*} \frac{q}{r} e^{-\beta r}
\end{equation}
\noindent yields an electric field, ${\mathbf E}_Y=-\nabla \Psi_Y$. 
Since $\nabla$ in spherical coordinates reads 
\begin{equation}
\nabla =\frac{\partial}{\partial r} {\mathbf u}_r+\frac{1}{r} \frac{\partial}{\partial \theta} {\mathbf u}_{\theta}+\frac{1}{r \sin\theta} \frac{\partial}{\partial \phi} {\mathbf u}_{\phi}, 
\end{equation}
\noindent then, 
\begin{eqnarray}
\label{eq07}
{\mathbf E}_Y &=&-\frac{\partial \Psi_Y}{\partial r} {\mathbf u}_r \nonumber\\ 
\, &=&-\frac{\partial \Psi_Y}{\partial r}  \left( \cos\theta {\mathbf u}_z+\sin\theta \cos\phi {\mathbf u}_x+\sin\theta \sin\phi {\mathbf u}_y\right) ,
\end{eqnarray}
%
%\begin{multline}
%\label{eq07}
%{\mathbf E}_Y=-\frac{\partial V_Y}{\partial r} {\mathbf u}_r = -\frac{\partial V_Y}{\partial r}\times \\ 
%\left( \cos\theta {\mathbf u}_z+\sin\theta \cos\phi {\mathbf u}_x+\sin\theta \sin\phi {\mathbf u}_y\right),
%\end{multline}
%
%
\noindent and the $z$-component of the field is: 
\begin{equation}
\label{eq08}
E_z =-\frac{\partial \Psi_Y}{\partial r} \cos\theta = \frac{q}{\epsilon_1^*} \frac{(1+\beta r) e^{-\beta r}}{r^2} \cos \theta.
\end{equation}
The boundary conditions at the interface between this medium and a rigid (i.e., non-polaronic) one defined by a static dielectric constant $\epsilon_2$ are given by Eq.~(\ref{eq06}). %, where $\sigma$ is the induced charge density at the interface. 
 Then,
\begin{equation}
\label{eq09}
\epsilon_1^* (E_z-\frac{\sigma}{2 \epsilon_0})=\epsilon_2 (E_z+\frac{\sigma}{2 \epsilon_0})  \to \frac{\sigma}{ \epsilon_0}= 2 E_z \frac{\epsilon_1^*-\epsilon_2}{\epsilon_1^*+\epsilon_2}.
\end{equation}
Using Eq.~(\ref{eq08}) we get:
\begin{equation}
\label{eq10}
\sigma = \frac{1}{2\pi}\frac{q}{\epsilon_1^*} \frac{\epsilon_1^*-\epsilon_2}{\epsilon_1^*+\epsilon_2} \frac{f(r)}{r^2} \cos\theta,
\end{equation}

\noindent with $f(r)=(1+\beta r) e^{-\beta r}$. 
Note that this induced charge density is identical to that generated by a standard Coulomb potential in a rigid medium, but replacing 
$1/r^2$ by $f(r)/r^2$.\\
 
It should be noted that in the limit of zero polaron radius ($\beta \to \infty$), $(1+\beta r) e^{-\beta r} = 0$. 
Then, the Yukawa induced charge density is zero. 
In any other case, we can replace the effect that this induced charge produces by the effect of an image charge $q_{i}$ 
(located either in the position of the original charge or in its mirror position, 
i.e., always located in the opposite region to that on which it acts) of magnitude
\begin{equation}
	q_i = \frac{q}{\epsilon_1^*}\,\frac{\epsilon_1^*-\epsilon_2}{\epsilon_1^*+\epsilon_2},
\end{equation}
\noindent which generates a Yukawa potential $q_{i} e^{-\beta r} / ( 4 \pi \epsilon_0\, r)$ 
instead of a Coulomb $q_{i} / (4 \pi \epsilon_0 \, r)$ one.\\

In summary, except that $1/r^2$ must be multiplied by a $f(r)$ term, the image charges are calculated as in standard rigid media. 
Consequently, the potential between electrons and holes in layered metal halide perovskites 
(considering both, dielectric confinement and polaronic contributions) can be obtained by slightly modifying the expressions 
provided in Ref. \cite{KumagaiPRB} for rigid semiconductor quantum well-like structures. 
The modified potential reads as follows (back in a.u.):
\begin{equation}
\label{eqVeh}
V_{eh}=-\sum_{k=C,Y} \; \sum_{n=-\infty}^{\infty}\frac{q^{2}\,q_{n,k}\,f_{n,k}({\mathbf r}_e,{\mathbf r}_h)}{r_n}.
\end{equation}
\noindent Here,
\begin{equation}
\label{eqrn}
r_n = \left \{ ({\mathbf r}_{\parallel ,e}-{\mathbf r}_{\parallel ,h})^2 + \left [ z_e - (-1)^n z_h + nL_z \right ]^2 \right \}^{1/2},
\end{equation}
\noindent with $L_z$ the thickness of the perovskite layers, and
\begin{equation}
	q_{n,k} = \frac{1}{\epsilon_k}\, \left [ \frac{\epsilon_k - \epsilon_2}{\epsilon_k - \epsilon_2} \right ]^{\left | n \right |},
\end{equation}
\noindent  with $\epsilon_C = \epsilon_s$ (the static dielectric constant of the perovskite material), 
$\epsilon_Y = \epsilon^*$ (ionic correction of the perovskite), and $\epsilon_2$ the dielectric constant of the organic ligands.
The functions $f_{n,k}$ are given by:
\begin{eqnarray}
\label{eqf}
	f_{n,C}({\mathbf r}_e,{\mathbf r}_h) &=& 1, \\
	f_{n,Y}({\mathbf r}_e,{\mathbf r}_h) &=& \frac{1}{2} \left ( e^{-\beta_e r_n} + e^{-\beta_h r_n} \right ).
\end{eqnarray}

%\noindent where $r_n = \{ ({\mathbf r}_{\parallel ,e}-{\mathbf r}_{\parallel ,h})^2 + [ z_e - (-1)^n z_h + nL_z ]^2 \}^{1/2}$.

 The self-polarization potentials of electron and hole can be obtained from eqn (\ref{eqVeh}) by setting ${\mathbf r}_e = {\mathbf r}_h$, eliminating the $n=0$ terms (which represent the interaction of a charge carrier with itself), and dividing by 2 as corresponds to a self-energy:\cite{Jackson_book}
\begin{equation}
\label{eqself}
V_{i=e,h}^{self} = \sum_{k=C,Y} \; \sum_{n=\pm 1, \pm 2, \dots} \frac{q^{2}\,q_{n,k}\,g_{n,k}(z_i)}{2 \left | z_i - (-1)^n z_i + n L_z \right |},
\end{equation}
\noindent where now 
\begin{eqnarray}
\label{eqg}
	g_{n,C}(z_i) &=& 1 , \\
	g_{n,Y}(z_i) &=&  e^{-\beta_i  \left | z_i - (-1)^n z_i + n L_z \right |}.
\end{eqnarray}

Once Eqs.~(\ref{eqself}) and (\ref{eqVeh}) are plugged into Hamiltonian (\ref{eqH}), 
the exciton ground state is obtained by integrating it with a variational wavefunction of the form
\begin{equation}
\label{eqtrial}
\psi_X=N \, \Phi_e \Phi_h e^{-\frac{\sqrt{({\mathbf r}_{\parallel, e} - {\mathbf r}_{\parallel, h})^2}}{a_B^*}},
\end{equation}

\noindent where $N$ is a normalization factor, while $\Phi_e$ and $\Phi_h$ are the electron and hole single particle states, 
which account for quantum confinement. The exponential term is an in-plane Slater correlation factor, 
which captures excitonic interactions.  $a_B^*$ is the effective Bohr radius, which is optimized variationally 
by following the computational scheme described in Refs. \cite{PlanellesTCA} and \cite{RajadellPRB}, including the
modified potentials $V_{eh}$ and $V_i^{self}$ to include the Yukawa-type potentials. 
At this regard, and for the sake of computational feasibility in the calculation of $V_{eh}$, 
the exponential factor in the Yukawa potential is spanned as a sum of five gaussian functions,
including five linear and five non-linear fitting parameters, 
which are optimized from the starting set provided in Ref. \cite{StewartJCP}. 

A Mathematica code to carry out the above described procedure is provided in the Electronic Supporting Information for free use. The code has minimal memory requirements and can be executed in ordinary computers, providing exciton total and binding energies (sub-meV converged), effective Bohr radius and electron-hole overlap integral.

\section{Results and Discussion}
\label{secresults}

In this section, our model is used to investigate the connection between polaronic, dielectric confinement
and quantum confinement effects in quasi-2D halide perovskite layers.
For illustrative purposes, we consider MAPbI$_3$ (MA=methylammonium). 
Effective parameters for this material have been proposed in the literature: 
dielectric constants $\epsilon_{s1} = 22.0$ and $\epsilon_{\infty 1} = 5.6$,\cite{SaporiNS} 
effective masses $m_e=0.19$, $m_h = 0.22$, and phonon frequency $\hbar \omega_{LO}=16.5 meV$,\cite{BaranowskiAEM} 
which yield polaron radii of $l_e = 1/\beta_e = 0.94$ nm and $l_h = 1/\beta_h = 1.01$ nm. 
 %For simplicity, we use the mean polaron radius, $l=(l_e+l_h)/2$.
Unless otherwise noted, the low polarizability of the organic environment is characterized by 
$\epsilon_2 = 2$.\cite{BlanconNC} % (static and high frequency constants are generally similar in these molecules\cite{SaporiNS}).\\

\begin{figure}[h]
\centering
\resizebox{0.6\columnwidth}{!}{\includegraphics{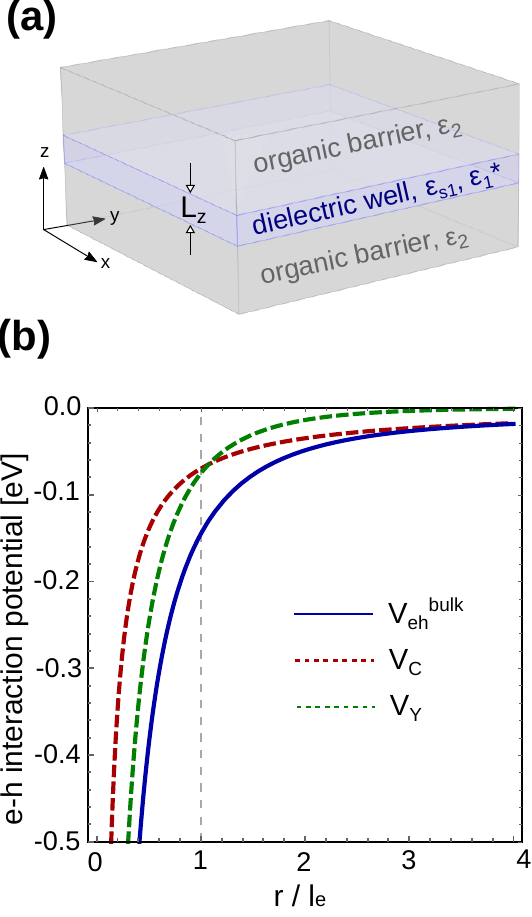}}
\caption{(a) Schematic of the system under study.
	(b) Haken potential (blue, solid line) in bulk MAPbI$_3$ as a function of the electron-hole distance in units of the polaron radius $l_e$. Coulomb ($V_C$, red dashed line) and Yukawa ($V_Y$, green dashed line) contributions are also included. Note that these contributions invert their predominance at radii close to the polaron radius (highlighted by a vertical, dashed line).}
\label{fig1}
\end{figure}
 
Figure \ref{fig1}(a) shows the different material domains of the system under study. 
The perovskite layers are modeled as a dielectric NPL sandwiched by organic barriers,
with abrupt (step function) interfaces.
The lateral dimensions of the NPL are fixed at 30x30 nm$^2$. That is in the weak confinement
(quantum-well like) regime. The number of layers defining the NPL thickness, $n_L$, is a variable.  
Each layer is constituted by a single octahedron (unit cell) with lattice parameter $a_0=0.63$ nm,\cite{FerreiraPRL} 
which add up to give a NPL thickness of $L_z=n_L\,a_0$. 

Prior to dealing with the numerical simulations for heterostructures, it is instructive to get a qualitative view
on the Haken potential in bulk MAPbI$_3$, as it defines the basic mechanism of carrier-lattice coupling we aim to study.
Figure \ref{fig1}(b) shows the profile of $V_{eh}^{bulk}$, given by Eq.~(\ref{eq02}),
 as a function of the electron-hole distance, in units of the electron polaron radius $l_e$ (blue line). 
Coulomb ($V_C$, red line) and Yukawa ($V_Y$, green line) contributions are also included. 
As can be observed, the Yukawa term is negligible for e-h separations of the order of 4$l_e$, 
where $V_{eh}^{bulk} \approx V_C$. %the Haken potential is roughly described by the Coulomb contribution. 
However, at lower e-h distances both contributions become significant.
Eventually, for $r \leq l_e$, $V_Y$ becomes more important than $V_C$.
In short, the smaller the exciton Bohr radius, the more relevant the ionic corrections.
Major changes are expected when it approaches the polaron radius. 

\begin{figure}[h]
\centering
\resizebox{0.5\columnwidth}{!}{\includegraphics{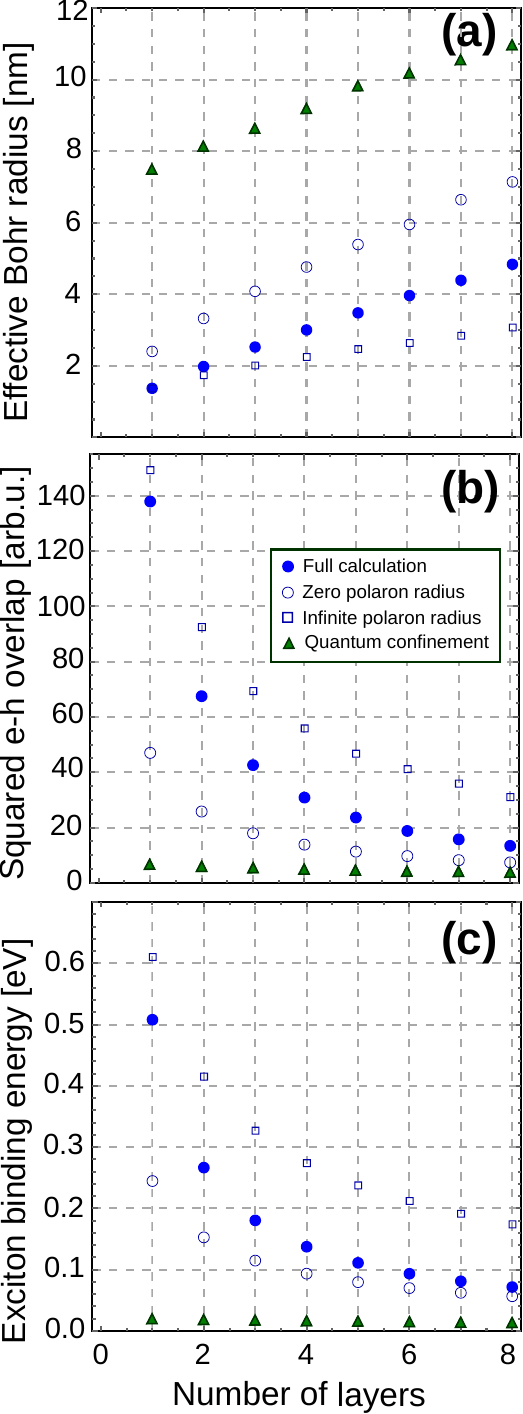}}
\caption{Thickness dependence of exciton properties in MAPbI$_3$ layers. 
	(a) Exciton effective Bohr radius. 
	(b) Electron-hole overlap squared. 
	(c) Exciton binding energy. 
	The organic barriers have $\epsilon_2=2$. 
	Results are depicted for different degrees of approximation in the model. 
	Polaron and dielectric mismatch contributions (full dots); limits of zero (empty circles) and infinite (squares) polaron radii; and absence of both, polaronic and dielectric confinement contributions (triangles).}
\label{fig2}
\end{figure}

We shall see next that quantum and dielectric confinement greatly reduce the exciton radius in layered 
perovskites with respect to their bulk value, and this indeed has a pronounced impact on its properties.
%
%Green triangles in \ref{fig2}(a) represent the calculated exciton Bohr radius $a_B^*$ (our variational parameter)
%in the absence of dielectric confinement and polaronic contributions. One can see the Bohr radius is
%reduced by quantum confinement. A much greater decrease is however obtained when dielectric confinement
%
Full dots in Fig. \ref{fig2}(a) represent the calculated exciton Bohr radius $a_B^*$ (our variational parameter) 
as a function of the number of layers $n_L$ conforming the NPL. For comparison, results excluding 
dielectric and polaronic contributions are also shown in the figure (green triangles; $\epsilon_{s1}$ is assumed). 
It is clear from the trend of the triangles that the exciton radius diminishes with respect to its bulk value
when the NPL becomes thinner. This is because the vertical quantum confinement reinforces exciton interactions in the plane.\cite{RajadellPRB}
Yet, the reduction is about 4 times greater in the presence of dielectric confinement and polaronic corrections (full dots).
The overall result is a surprisingly small exciton Bohr radius, which, for low enough $n_L$ is 
comparable to that obtained by considering infinite polaron radii (open squares in the figure),
i.e. as if $V_{eh}=-q^2/(\epsilon_{\infty 1} r)$.
In other words, in the few-layer limit the exciton becomes immune to the high ionic polarizability of metal halide perovskites.\\

Such small Bohr radii are expected to cause a sharp increase in the radiative recombination rates. The rate is proportional to the squared electron-hole overlap function, $S_{eh}^2=| \langle \psi_X | \delta_{r=0} | 0 \rangle |^2$, which can be calculated analytically within our model.\cite{RajadellPRB}  
Fig.~\ref{fig2}(b) confirms the rapid increase of $S_{eh}^2$ with decreasing number of perovskite layers. 
This is consistent with the experimental observation of reduced photoluminescence lifetimes as the perovskite NPLs become thinner,\cite{HintermayrAM}
although a more direct comparison is hindered by the co-existence of non-radiative processes.
Fig.~\ref{fig2}(b) also shows that polaronic effects raise $S_{eh}^2$ up to three times as compared to the case of 
static dielectric constant (cf full dots and empty circles). This suggests that polaronic effects in 2D layered perovskites
enable faster radiative recombination than in more rigid semiconductors, such as metal chalcogenide NPLs.

Dielectric, polaronic and quantum confinement contributions also affect noticeably other exciton properties of interest, 
namely its binding energy $E_b$. This is shown in Fig.~\ref{fig2}(c).  
Because the exciton Bohr radius in a strictly 2D system is half that of the 3D case, and $E_b = \hbar^2/(2\,\mu\,a_B^{*2})$ (with $\mu$ the exciton reduced mass), the relationship $E_b^{2D} = 4 E_b^{3D}$ has been often cited in the literature to summarize the
effect of quantum confinement.\cite{ChakraPCCP,HintermayrAM}
A close inspection to the green triangles in Fig.~\ref{fig2}(c) confirms that $E_b$ indeed increases as the 
perovksite becomes thinner by action of quantum confinement.
However, the scale of the increase is greatly magnified upon inclusion of dielectric and polaronic terms (full dots).
It is only in this case that the order of the binding energies ($E_b=100-500$ meV) is in agreement with that 
measured in related structures ((BA)$_2$(MA)$_{n-1}$Pb$_n$I$_{3n+1}$ perovskites with 1-5 layers, Ref.~\cite{BlanconNC}).

We can disentangle dielectric and polaronic contributions in the analysis of $E_b$. 
The impact of dielectric confinement can be measured by the difference between green triangles 
 and open circles in the figure. The former include only quantum confinement, while the latter
 add dielectric confinement but no polaronic effects (static dielectric constant).
 It is clear that dielectric confinement alone suffices to produce a large effect on $E_b$,
 from a 4-fold increase when $n_L=8$ to a 14-fold increase when $n_L=1$.
 This can be explained as a consequence of the large polarizability of metal halide perovskites. 
 On the one hand, the large $\epsilon_{s1}$ constant of the perovskite makes the bare Coulomb contribution small.
 This justifies the small $E_b$ values of the green triangles.
 On the other hand, the large $\epsilon_{s1}/\epsilon_2$ ratio implies a large dielectric contrast with the 
 organic medium, which leads to enhanced surface polarization charges (or the image charges they project).
 These intensify the excitonic interaction, which results in much larger values of $E_b$ for the empty circles.

 The effect of polaronic contributions can be visualized by comparing empty circles and full dots in Fig.~\ref{fig2}(c).
 They become increasingly noticeable when the number of layers decreases, because $a_B^*$ becomes comparable to the polaron
 radius (recall Fig.~\ref{fig2}(a)). For $n_L=1$, the enhancement exceeds a factor of 2 (i.e. 100\% increase).
 It is worth noting too that even for thin NPLs, the actual $E_b$ we calculate (full dots) have not reached the
 values expected in the limit of infinite polaron radius (squares). This means that relationships of the type
 $E_b^{2D} \propto  (\epsilon_{\infty 1} / \epsilon_2) $, which are sometimes taken as reference in the literature,\cite{ChakraPCCP}
 provide an upperbound estimate.

Taken together, it is remarkable that the exciton binding energies are comparable or greater to those of II-VI semiconductor 
NPLs,\cite{RajadellPRB, ShornikovaNL, ZelewskiJPCL, YangJPCc, AyariNS, MaciasJPCc} despite the large polarizability of perovskites.
For $n_L \leq 3$, the joint effect of dielectric and polaronic contributions yields binding energies of the order of several hundreds of meV, much larger than those expected by omitting polaronic effects and in agreement with reported experimental and theoretical results.\cite{SichertNL, ChakraPCCP, BlanconNC, TanakaPRB}\\

\begin{figure}
\centering
\resizebox{1.0\columnwidth}{!}{\includegraphics{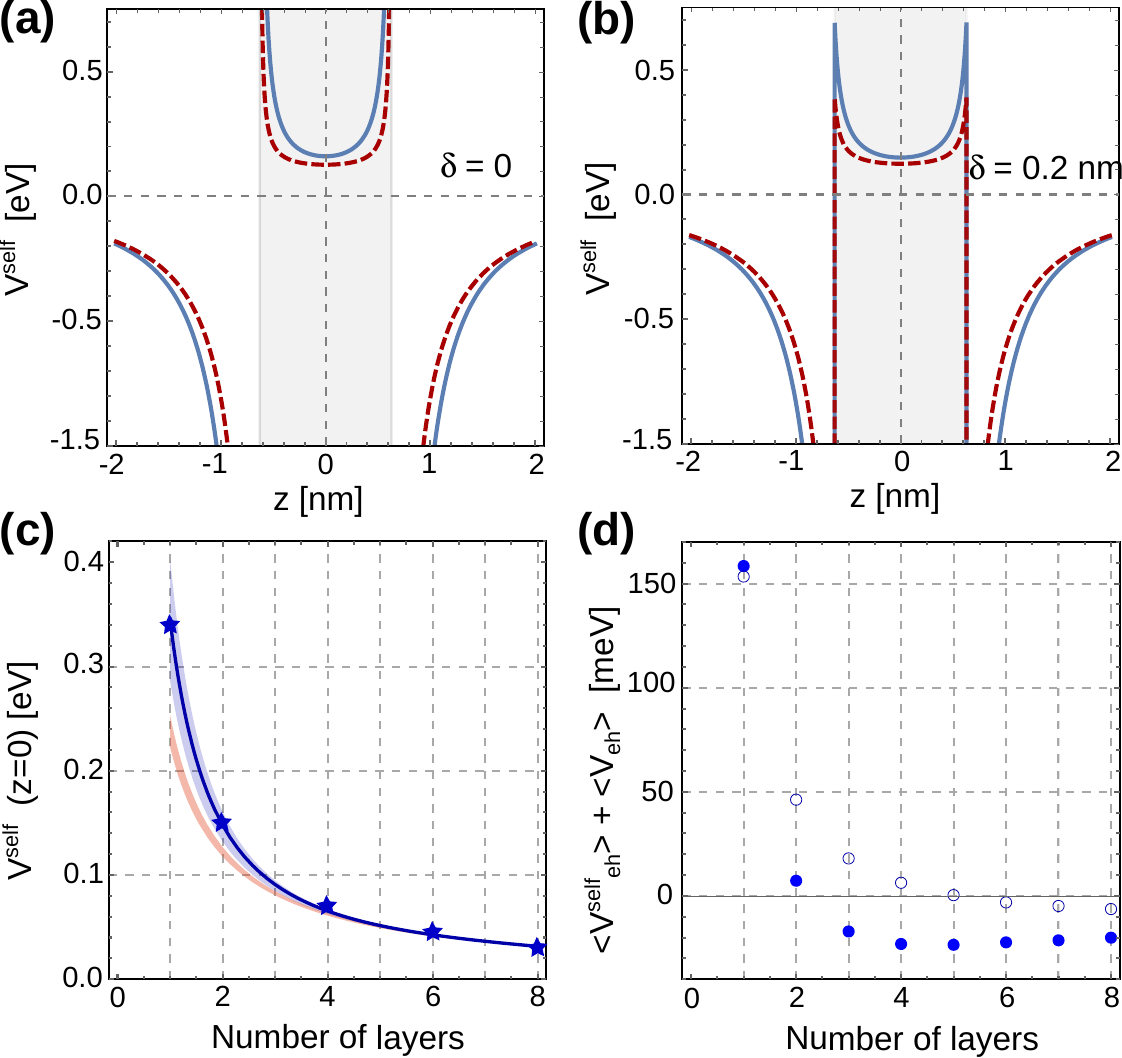}}
	\caption{ (a-b) Cross-section of the self-polarization potential along the strong confinement axis of a two-layer 
	MAPbI$_3$ quantum well, with (blue line) and without (red line) the Yukawa correction term. 
	The external medium has $\epsilon_2 = 1$: 
	(a) The method of shifted-mirror faces is not taken into account ($\delta = 0$); 
	(b) $\delta = 0.2$ nm. 
	(c) Self-polarization potential at the center of the MAPbI$_3$ quantum well structure as a function of the number of layers. 
	Stars: DFT data from Ref. \cite{SaporiNS} for (BA)$_2$(MA)$_{n-1}$Pb$_n$I$_{3n+1}$. 
	Blue line: Results obtained for $\delta = 0.2$ nm. 
	Blue-shaded region: results for $\delta=0-0.6$  nm.
	Orange-shaded region: same but disregarding polaronic contributions.
	(d) Sum of the electrostatic interaction energies 
%	(self-energies $\Sigma_{eh}$ and electron-hole interaction energy $E_{eh}$) 
	for the system considered in Fig. \ref{fig2} as a function of the number of layers in the structure. 
	Full dots (empty circles) include (exclude) polaronic contributions.} 
\label{fig3}
\end{figure}

The exciton properties presented so far are barely affected by the self-polarization potentials. %, $V^{self}_e$ and $V^{self}_h$. 
However, a complete description of dielectric confinement requires their inclusion. 
 The weak dielectric screening of the organic barriers enhances not only the e-h Coulomb interaction ($V_{eh}$), 
but also the interaction of carriers with their own image charges ($V^{self}_e$ and $V^{self}_h$).
The two effects are known to nearly compensate each other in small crystallites.\cite{BrusJCP, BolcattoPRB} 
However, in strongly anisotropic systems such as colloidal II-VI nanoplatelets, they have been shown to render 
sizable global effects on their emission wavelengths or composite properties.\cite{PolovitsynCM, MovillaJPCL} 
 Further evidence in this direction has been recently obtained in Cs-Pb-Br composites.\cite{CaicedoAFM}
Below we show that similar effects take place in layered perovskites, and that our image charge method gives
a description consistent with atomistic (DFT) simulations.

Figure \ref{fig3}(a) shows the $z$-cross section of the self-polarization potential, $V^{self}$, 
calculated from Eq.~(\ref{eqself}) for a 2-layer MAPbI$_3$ system surrounded by air or vacuum ($\epsilon_{2} = 1)$. 
 The profile of $V^{self}$ is similar with (blue line) and without (red line) polaronic interactions.
For carriers confined within the NPL, $V^{self}$ acts as a repulsive potential, which becomes attractive outside.
The effect of polaronic interactions is to provide a moderate increase of the repulsions and attractions (notice
the shift between lines).
 
On the dielectric interfaces, the potential yields non-integrable divergences. 
This is a well-known limitation of the image charge method when abrupt dielectric interfaces are assumed,\cite{KatanCR}
arising from the coincident position of carriers and their image charge.
 Several strategies have been proposed to circumvent this non-physical result.\cite{BanyaiPRB, BanyaiBOOK, BolcattoJPCM, MovillaCPC10}
 All of them rely on the argument that the notion of dielectric constant loses its meaning at distances 
 comparable to the interatomic one, making the abrupt, stepwise dielectric profile model an oversimplification of the dielectric interface. 

One strategy to sort this problem is the so-called method of the shifted mirror faces.\cite{KumagaiPRB} 
It consists in introducing a phenomenological parameter $\delta$, physically related to the thickness of a 
finite-width interfacial layer between adjacent materials. 
 Mathematically, the method simply replaces the $n=\pm 1$ terms in Eq.~(\ref{eqself}) with the following:
\begin{align}
\label{eqn1terms}
&\frac{q^{2}\, q_{1,C}}{2 \left | 2 z_i + L_z + \delta \right |} + \frac{q^{2}\, q_{1,Y}\, e^{-\beta_i  \left | 2 z_i + L_z + \delta \right |}}{2 \left | 2 z_i + L_z + \delta \right |}\nonumber\\
+ &\frac{q^{2}\, q_{-1,C}}{2 \left | 2 z_i - L_z - \delta \right |} + \frac{q^{2}\, q_{-1,Y}\, e^{-\beta_i  \left | 2 z_i - L_z - \delta \right |}}{2 \left | 2 z_i - L_z - \delta \right |},
\end{align}
\noindent with $\delta$ in the range of a few Angstroms.
Figure \ref{fig3}(b) displays the same case as in Fig. \ref{fig3}(a) 
but employing the method of the shifted mirror faces with $\delta = 0.2$ nm. 
Now, the self-potential is still discontinuous, yet integrable. 
It is worth noting that its value influences $V^{self}$ not only at the interfaces, 
but also throughout all space. Then, the shift parameter $\delta$ turns into a fitting parameter 
--the only one-- of this empirical model.

Because quantum confinement sets the peak of the exciton charge density in the $z=0$ plane of the NPL,
the value of the self-polarization potential on this plane is particularly important in determining the exciton energy, 
and hence the optical band gap.
In Fig. \ref{fig3}(c), $V^{self}(z=0)$ is represented as a function of the number of perovskite layers. 
The blue-shaded region delimits the results obtained by our model for $\delta$ values ranging from 0 to 0.6 nm 
(upper and lower limits, respectively). 
For comparison, the results obtained in Ref. \cite{SaporiNS} by means of atomistic (DFT) calculations are also included (blue stars). 
As can be seen, our model compares very well with the DFT results for a shift parameter of $\delta = 0.2$ nm (blue line in the figure). 
This agreement not only supports the validity of our model, but it also highlights the need to account for the non-abrupt
character of dielectric interfaces (i.e. finite $\delta$) in these materials, especially for small $n_L$, 
a fact which had been independently pointed out by DFT calculations.\cite{KatanCR,TraoreACS}
 It is also worth stressing that the agreement with atomistic methods cannot be reached without including 
 polaronic ($V_Y$) effects in the model. To make this point clear, in Fig.~\ref{fig3}(c) we also show the self-energy
 obtained for the same range of $\delta$ values, but taking into account $V_C$ contributions only (orange-shaded region).\\

After integrating Eq.~(\ref{eqH}) on the variational function $\Psi_X$, 
the exciton energy can written as:
\begin{equation}
	E_X = \langle E_{conf} \rangle + \langle V^{self}_{eh} \rangle + \langle V_{eh} \rangle + E_{gap}^\Gamma,
\end{equation}
where $\langle E_{conf} \rangle$ is the quantum confinement energy, $\langle V^{self}_{eh} \rangle$ 
the sum of electron and hole self-energy repulsions and $\langle V_{eh} \rangle$ the Coulomb attraction energy.
As mentioned before, dielectric confinement enhances both $\langle V^{self}_{eh} \rangle$ and $\langle V_{eh} \rangle$.
We study the compensation between the two terms in Fig.~\ref{fig3}(d), which is relevant in determining the total exciton energy.

%Thus, it is of interest to explore the role of the Yukawa interaction in this regard.
Fig.~\ref{fig3}(d) shows the totality of the electrostatic contributions 
($\langle V^{self}_{eh} \rangle + \langle V_{eh} \rangle$), versus the number of layers of our MAPbI$_3$ NPL. 
Full dots (empty circles) are simulations including (excluding) polaronic terms.
In all instances, the lack of compensation between self-energy and Coulomb polarization is evident. %and clearly reflects this lack of compensation.
As a matter of fact, for thin NPLs ($n_L \leq 2$), the energetic destabilization caused by $\langle V^{self}_{eh} \rangle$ 
is greater than the stabilization caused by $\langle V_{eh} \rangle$, such that the sum becomes positive.
This means that thin perovskite layers should experience an electrostatic blueshift in the presence of dielectric confinement,
despite the large exciton binding energy.
Polaronic interactions do not alter this trend qualitatively.\\

\section{Concluding remarks}

We have presented a model to describe exciton states in quasi-2D layered halide perovskites.
 The model can be used to disentangle the roles of quantum confinement 
(described through effective mass theory), dielectric confinement (image charge method)
and polaronic effects (Haken-like potential) in these materials,
which facilitates their rational design to achieve desired optoelectronic properties.

We have used the model to study layered MAPbI$_3$ structures. The results show that
polaronic effects become increasingly important as the number of layers decreases.
This is because quantum and dielectric confinement reduce the exciton radius, 
making it similar to that of the polaron. 
 It follows that excitons in few-layer structures are largely insensitive to the 
ionic polarization of the lattice, their response being close to that expected 
from the high frequency dielectric constant.

All in all, our work paves the way for the use of effective mass based models in the 
description of exciton properties of layered metal halide perovskites. 
Similar to the case of II-VI nanoplatelets, these can be expected 
to complement ab-initio methods with computationally affordable, 
yet intuitive and reliable descriptions of the optoelectronic properties. 
Such a possibility can be particularly helpful in the field of hybrid metal halide perovskite nanostructures, 
where atomistic simulations of excitons become extremely demanding,
 due to the large unit cells of these materials, the strong spin-orbit interaction (Pb), 
lack of translational symmetry, varying orientation of organic countercations, 
and the requirement of accounting for strong e-h correlations (e.g. through Bethe-Salpeter method).\cite{ChakraPCCP}

Computational codes associated with our model are provided along with this work (ESI),
and can be readily used to investigate excitons in layered perovskites built of different materials,
by simply providing input bulk effective masses, dielectric constants and LO phonon frequency,
which are often found in the literature, along with the dimensions of the structure.

%\section*{Author Contributions}
%JIC conceived the study.  JP and JLM developed the theoretical model (methodology). 
%JLM implemented the computational code (software) and ran the simulations (data curation).
%All the authors contributed to the analysis and discussion of results,
%as well as to the writing of the manuscript.

%\section*{Conflicts of interest}
%The authors declare no competing financial interests.

\begin{acknowledgments}
We acknowledge support from Grant No. PID2021-128659NB-I00, funded by Ministerio de Ciencia e Innovaci\'{o}n (MCIN/AEI/10.13039/501100011033 and ERDF A way of making Europe) and Generalitat Valenciana Prometeo Project No. 22I235-CIPROM/2021/078.
\end{acknowledgments}

%%%REFERENCES%%%
\bibliographystyle{apsrev4-2}
\bibliography{bib_pv_arxiv}

\end{document}